\documentclass[nologo,11pt,a4paper]{ETHpaper}

\usepackage[font=small,labelfont=bf,justification=centerlast]{caption}

\usepackage[numbers,compress]{natbib}

\usepackage{mathtools,amsmath,amssymb}
\DeclarePairedDelimiter{\ceil}{\lceil}{\rceil}
\usepackage[inline]{enumitem}
\usepackage{amsthm}

\usepackage{dcolumn}%
\usepackage{bm}%

\newcommand{\mean}[1]{\left\langle #1 \right\rangle}
\newcommand{\abs}[1]{\left| #1 \right|}

\title{Consensus from group interactions: An adaptive voter model on hypergraphs}
\titlealternative{Consensus from group interactions: An adaptive voter model on hypergraphs}
\author{Nikos Papanikolaou$^{1}$, {Giacomo Vaccario$^{1}$}, {Erik Hormann$^{2}$},\\ {Renaud Lambiotte$^{2,3}$}, {Frank Schweitzer$^{1,3,\star}$}}

\authoralternative{N. Papanikolaou, G. Vaccario, E. Hormann, R. Lambiotte, F. Schweitzer}

\address{$^{1}$Chair of Systems Design, ETH Zurich, Weinbergstrasse 58, 8092 Zurich, Switzerland \\
{$^{2}$Mathematical Institute, University of Oxford, 
Woodstock Road, Oxford
OX2 6GG, UK} \\
{$^{3}$Complexity Science Hub Vienna, Josefstädterstrasse 39, 1090 Vienna, Austria}

\footnotetext{$^{\star}$corresponding author:\ {fschweitzer@ethz.ch}}
  \url{www.sg.ethz.ch}}
\reference{\em (Submitted for publication)} 
\makeframing

\begin{document}

\maketitle

\begin{abstract}
  We study the effect of group interactions on the emergence of consensus
  in a spin system.
  Agents with discrete opinions $\{0,1\}$ form groups.
  They can change their opinion based on their group's influence (voter dynamics), but groups can also split and merge (adaptation). 
  In a hypergraph, these groups are represented by hyperedges of different sizes.
  The heterogeneity of group sizes is controlled by a parameter $\beta$. 
  To study the impact of $\beta$ on reaching consensus, we provide extensive computer simulations  and compare them with an analytic approach for the dynamics of the average magnetization.
  We find that group interactions amplify small initial opinion biases, accelerate the formation of consensus and lead to a drift of the average magnetization. 
  The conservation of the initial magnetization, known for basic voter models, is no longer obtained. 
\end{abstract}

\section{Introduction}
The formal analysis of spin systems in statistical physics dates back to 1920, when Wilhelm Lenz and Ernst Ising developed the first mathematical model of ferromagnetism \citep{niss2005history}.
Today, we could rightly call it a multi-agent model \citep{krasnytska2020ising,Schweitzer2018}.
The entities, i.e. the atoms or agents, are characterized by a degree of freedom, their spin $s_{i}(t)\in \{-1,+1\}$, which is a discrete variable representing the direction of the magnetic dipole moment.
The $s_{i}(t)$ can change over time because of interactions between agents, which are expressed by coupling constants $J_{ij}$.
Whether or not two agents $i$ and $j$ can interact is defined by an underlying topology, for instance, a one or two-dimensional regular lattice in the case of the original Ising-Lenz model. 
To calculate the macroscopic state, it is assumed that the dynamics can be {decomposed} into \emph{pairwise interactions} between any two agents $i$ and $j$, with $J_{ij}=0$ if these agents are not neighbors.

In this paper, we extend this view by considering additional \emph{group interactions}.
A group refers to a number of agents which interact jointly and simultaneously, i.e. the group interaction {cannot} be decomposed into pairwise interactions.
Such group interactions are relevant in diverse systems composed of many agents,
ranging from  physics \cite{goban2018emergence,matheny2019exotic,franz2001ferromagnet} to neural networks \citep{memmesheimer2012non} and ecology \cite{levine2017beyond}.  
To adequately represent group interactions,  
in Sect.~\ref{sec:hypergraph} we utilize the concept of a \emph{hypergraph}~\citep{battiston2020networks}. 
Like a network, a hypergraph consists of nodes and edges. %
In a simple network these edges represent pairwise interactions, while in a hypergraph \emph{hyperedges} represent groups of agents.
Hence, hyperedges can have different sizes and 
interactions between different groups are represented by overlaps between hyperedges.

To define interactions between groups of agents, we use  the \emph{voter model}, a well-studied spin system \citep{sood2008voter,castellano2005comparison,suchecki2005conservation,lb-fs-03}. 
The term ``voter''  \citep{fernandez2014voter,redner2019reality} refers to an agent with a discrete ``opinion'', $s_{i}(t)\in \{0,1\}$, i.e. the agent is either in favor ($s_{i}=1$) or against ($s_{i}=0$) a given issue.
In the most basic version of the voter model, two agents $i$ and $j$ are randomly chosen from the whole system, and agent $j$ gets assigned the opinion of agent $i$, i.e. $s_{j}(t+1)=s_{i}(t)$.
This means, if the two agents already had the same opinion, nothing changes; otherwise, the opinion of agent $i$ is replicated.
Extended versions of the voter model consider specific network topologies, such as lattices or complex networks, to define the pairs of agents that can interact \citep{suchecki2005voter,castellano2003incomplete,zschaler2012early}.
The main question always regards the final macroscopic state: will all agents over time have the same opinion or not?
To quantify the outcome, the global fraction $f(t)$ of agents with opinion 1 is used, from which one can derive the \emph{magnetization} of the spin system, $M(t)=2f(t)-1$.
If either $M=+1$ or $M=-1$ in the long run, the final state is denoted as consensus. Otherwise one finds a coexistence of the two opinions.

For different variants of the  voter model, the time to reach consensus~\citep{krapivsky2010kinetic, stark2008decelerating}, the conservation of magnetization~\citep{liggett1985interacting, suchecki2005conservation}
and the coexistence of opinions~\citep{frachebourg1996exact, castellano2003incomplete, schweitzer2015neighborhood} have been studied.
To additionally consider the effect of group interactions,
the majority rule has been proposed \cite{krapivsky2003dynamics,noonan2021dynamics}.
It assumes that at each
timestep a number $n\geq 3$ of agents is selected and adopts the majority opinion in the group of selected agents.
In case of a stalemate, additional rules were proposed to break the symmetry \citep{galam2002minority}.

To relax the assumption of a static neighbourhood structure, which may be unrealistic in certain contexts, adaptive voter models have been proposed~\citep{holme2006nonequilibrium, durrett2012graph}.
They consider the co-evolution of agents' interactions and opinions~\citep{vazquez2008generic, gross2008adaptive}, often on different time scales.
We will build on these works with our modeling approach that applies adaptive voter models to hypergraphs. 
Our work complements, and contrasts, the work by \citet{horstmeyer2020adaptive} who also consider group interactions in the adaptive voter model, albeit on the basic level of  simplicial complexes of size two.
That means, group interactions are again decomposed, this time into interactions between three agents that form a triangle, or 2-simplex.
Additionally, pairwise interactions and nodes totally disconnected from others are considered.

In \citep{horstmeyer2020adaptive} the opinion dynamics inside 2-simplices is motivated by a peer pressure, following the majority rule.
That means, a majority of two agents (with the same opinion) influences a minority of one agent (with a different opinion) to adopt their opinion.
It should be noted that the influence of larger groups, and within larger groups, was already formalized in models of social impact \citep{lewenst-nowak-latane-92,hoiyst2000}. 
They even captured different forms of peer pressure, such as persuasion and support \citep{nowak-szam-latane-90}, taking social distances in larger groups and weights of opinions into account.
Also models of continuous opinion dynamics considered the weighted influence of larger in-groups \citep{Groeber2009,schweitzer-andres-2020-social}.

To overcome the shortcomings that result from the decomposition into 2-simplicies, we follow 
a route that was already mentioned, but not taken, in \citep{horstmeyer2020adaptive}, namely to build on hypergraphs.
These structures are already used to study the consensus dynamics resulting from the majority rule \citep{noonan2021dynamics}, but with a restriction to groups of size 3.
Here, we extend this approach to consider hypergraphs with groups of different and arbitrarily large sizes, $n$.
Their distribution $\pi(n)$ is controlled by a parameter $\beta$ that accounts for the \emph{heterogeneity} in group sizes.
We specifically study the impact of this heterogeneity on the consensus formation.

Our model extends the adaptive voter model by \citet{durrett2012graph} in that we generalize its rules for hypergraphs.
Specifically, in our model at each time step a group of size $n$ is selected.
If $n=2$, i.e. if we have pairwise interactions, we use the rules of opinion adoption and rewiring described in \cite{durrett2012graph}.
If $n\geq 3$, we propose a generalization of the adoption and rewiring processes, which we call influence and split-merge processes.
The former assumes that the minority in the group 
changes their opinion with a specific probability, the latter considers that the minority leaves the group and both majority and minority merge with two other groups.
An additional parameter $\gamma$ defines the threshold to distinguish between these processes and its impact is studied in our paper. 

We analyze this model using a variation of a mean field approximation called heterogeneous mean field (HMF) approximation where at each time step agents are randomly chosen to form a group with a size sampled from the  distribution $\pi(n)$.
This is a valid approximation for highly connected hypergraphs.
The focus of our paper is on the dynamics of the average magnetization.
We provide extensive computer simulations to study this dynamics dependent on the model parameters $\beta$, $\gamma$ and the system size.
Importantly, we also develop an analytical model with noise that can accurately describe the simulations for a wide range of parameters.
The analysis indicates that, compared to the case of only pairwise interactions, group interactions reduce noise, amplify small initial opinion biases and accelerate the convergence toward consensus.
At difference with the basic voter model, the initial magnetization of the system is not conserved.
Instead, we observe a drift of the average magnetization in the direction of the initial bias, which increases with the heterogeneity in group sizes. 

The remainder of this paper is divided into three sections.
In Section~\ref{sec:model}, we formally introduce hypergraphs and the dynamic assumptions of our model.
This is followed by three sections that present results at different levels.
Section~\ref{sec:results-comp-simul} contains results from computer simulations, to give an orientation about the dynamics and the impact of the different model parameters.
Section~\ref{sec:analytic-results} presents analytical results for the average magnetization without and with noise.
They allow to calculate a switching rate for the sign of the magnetization trajectories.
Section~\ref{sec:numerical-comparison} eventually compares analytical and simulation results numerically, to demonstrate their good agreement.
Finally, we present our conclusion in Sect.~\ref{sec:conclusion}.

\section{An adaptive voter model on hypergraphs}
\label{sec:model}
\subsection{The dynamics of the adaptive voter model} 
\label{sec:hypergraph}

In order to define our model, we remind that the following discussion addresses three different levels of model complexity.
On the basic level, we consider an adaptive voter model on a simple network, on the second level we generalize this model to hypergraphs and on the third level we apply a heterogeneous mean-field (HMF) approximation to the hypergraph model, in order to obtain analytical results.

As already mentioned, the basic voter model considers pairwise interactions between two agents $i$ and $j$.
These agents have to be neighbors to interact, i.e. if we assume ferromagnetic coupling the interactions have the form 
\begin{equation}
    J_{ij} =  
    \begin{cases}
        1 & \text{if }(i,j)\in E \\
        0 & \text{otherwise}
      \end{cases}
      \label{eq:1}
\end{equation}
where $E\subset V \times V$ denotes a set of edges between a set of nodes $V$, which both define a network $\mathcal{G}(V,E)$.
Specifically, the $J_{ij}$ defines the elements of the adjacency matrix that contains full information about the network topology.

Each agent $i$ is characterized by its spin, or state, or ``opinion'', $s_{i}(t)\in\{0,1\}$, which can change because of the interaction with neighboring agents.
As explained above, the basic dynamics reads:
\begin{align}
  \label{eq:2}
  s_{i}(t+1)=s_{j\in n_{i}}(t)
\end{align}
i.e. agent $i$ adopts the opinion of a randomly chosen agent $j$ from its neighborhood $n_{i}$.
If $i$ is connected to $n_{i}$ other agents on the simple network, i.e. it has  a degree $n_{i}$, then 
the probability that one of these neighbors has the opinion $s=1$ is
\begin{align}
  f^{(1)}_{i}(t)\equiv f_{i}(t)=\frac{1}{n_{i}}\sum_{j \in n_{i}} s_{j}(t)
  \label{eq:3}
\end{align}
where $f^{(s)}_{i}$ defines the local frequency of opinion $s$ in the neighborhood of $i$  \citep{fs-voter-03}.
We will normalize this to the opinion $s=1$, hence
$f^{(1)}_{i}=f_{i}$ , while $f^{(0)}_{i}=1-f_{i}$.
If $f_{i}(t)>0.5$, i.e. if opinion 1 is the majority opinion, then the opinion of $i$ in the next time step will be more likely also $s_{i}(t+1)=1$.
Specifically, $p_{i}(s_{i},t)$ denotes the probability to find agent $i$ with opinion $s_{i}$ at time $t$.
Its change in time is given by the master equation:
\begin{align}
  \label{eq:6}
  \begin{split}
&    p_{i}(s_{i},t+1) - p_{i}(s_{i},t)\approx \frac{dp_{i}(s_{i},t)}{dt}= \\ & w(1-s_{i}|s_{i}) p_{i}(s_{i},t) -  w(s_{i}|1-s_{i}) p_{i}(1-s_{i},t)
  \end{split}
\end{align}
where the transition rates in the case of the simple voter model read as \citep{fs-voter-03,lb-fs-03}
\begin{align}
  w_{i}(1-s_{i}|s_{i})\propto f_{i}^{1-s_{i}}
  \label{eq:7}
\end{align}
That means a change of the opinion of agent $i$ from $s_{i}$ to the opposite opinion $1-s_{i}$ linearly depends on the frequency $f^{1-s_{i}}$ of the opposite opinion in the neigborhood of $i$.
It is known that the linear voter model always reaches consensus if the underlying topology is connected.
But which opinion makes up for this consensus is randomly decided.

The systemic variable is the magnetization $M(t)$ which follows from the total fraction of agents with opinion 1:
\begin{align}
  \label{eq:5}
  M(t)=2 f(t) -1\;; \quad 
  f^{(1)}(t)\equiv f(t)=\frac{1}{N}\sum_{i}s_{i}(t)
\end{align}
It monitors whether the system will reach consensus, $\abs{M}=1$, or coexistence with $0\leq \abs{M}<1$.
The linear voter model shows a conservation of magnetization \citep{suchecki2005conservation}.
That means the initial fraction of agents with opinion 1, $f(0)$, already tells how often $T$ simulation runs will end up with a consensus of opinion 1 \citep{fs-voter-03}.

To obtain the system dynamics, i.e. the \emph{expected} dynamics of $M(t)$, we have to average over a large number of independent simulations, where
the initial condition $M(0)$ plays an important role.
Because we initially assign opinions randomly to agents, only the expected initial magnetization, $\mathbb{E}[M(0)]=\mean{M(0)}$, is fixed.
Actual values of $M(0)$ follow from a binomial distribution for $f(0)$.
So, we have two sources of randomness, or ``noise'', when initializing our system: (i) deviations of $M(0)$ from the expected value $\mean{M(0)}$, and (ii) deviations in initial configurations, i.e. specific assignments of opinions to agents, for the same value $M(0)$.
This noise can lead to different final states characterized by a different magnetization, as we will show below. 

In the following, we use the average $\mean{M(t)}$ to denote $\mathbb{E}[M(t)]$. 
An initial condition $\mean{M(0)}=0$ implies a symmetry between opinions.
Hence, averaging over many simulations would cancel out the dynamic effects that lead to consensus of one opinion.   
To avoid trivial outcomes, we therefore use a nontrivial initial condition $\mean{M(0)}> 0$, i.e. an initial bias that \emph{on average} breaks the symmetry in favor of opinion 1.
Note that specific initial configurations can still deviate from this. 
We will study the impact of this initial bias on the dynamics to reach consensus.

The \emph{adaptive} voter model adds a new degree of freedom to the dynamics.
Agents cannot only adapt their opinions, they can also rewire the links to their neighbors $j\in n_{i}$ with a certain probability $r$.
This applies only if the two agents $i$ and $j$ have different opinions.
Then, with a probability $1-r$, the dynamics follows the master Eqn.~\eqref{eq:6} for adaptation and their edge $(i,j)$ remains. 
With a probability $r$, this edge $(i,j)$ is deleted and two new edges $(i,k)$ and $(j,l)$ are created, where $s_{i}=s_{k}$ and $s_{j}=s_{l}$.
As a result, both $i$ and $j$ each are linked to an agent with the same opinion they have, and the density of the network, specifically the number of edges between agents with the same opinion, has increased.
The rewiring mechanism strengthens the respective majorities in the neighborhood of both agents $i$ and $j$. 

\subsection{From pairwise to group interactions}
\label{sec:dynamics}

Now we move the adaptive voter model to \emph{hypergraphs}. 
Similar to a network, a {hypergraph} $\mathcal{H}(V, \mathcal{E})$ is described by the set of nodes, or agents, $V$, and the set of \emph{hyperedges}, $\mathcal{E}$.
These differ from the edges $E$ of the simple network as they now connect \emph{groups} of agents.
Thus, hyperedges basically represent groups of different sizes.
A group ${a}$ is described by a $n$-tuple of agents, ${a}=(i,j,...)$, where $\abs{{a}}=n_{a}$ denotes the group \emph{size}.
That means, with $n_{a}$=2 we are back at the simple network that only considers pairwise interactions.
Therefore, we denote these edges \textit{simple-edges}.

We can adopt the above definitions for frequencies and transition rates on simple networks to now describe \emph{groups}.
The frequency of agents with opinion 1 in group $a$ follows from Eqn.~\eqref{eq:3}:
\begin{equation}
 f^{(1)}_{a}(t) \equiv f_{a}(t) = \frac{1}{n_{a}} \sum_{i \in {a}} s_i(t)\;;\quad f^{(0)}_{a}(t)=1-f_{a}(t)
  \label{eq:8}
\end{equation}
The majority of agents in group $a$ has opinion $s$ if $f^{(s)}_{a}>0.5$.
Further, the transition rate to change the opinion, Eqn.~\eqref{eq:7}, now refers to the group, i.e.
\begin{align}
  w(1-s_{i}|s_{i})\propto f_{a}^{(1-s_{i})}\;; \quad i \in {a}
  \label{eq:7a}
\end{align}
This implies that the group would not change any opinion, if all agents in the group have the same opinion.

While the adaptation of opinions inside the group does not differ from the dynamics assumed for the simple network, the rewiring mechanism becomes different if groups have a size $n_{a}>2$.
Instead of rewiring vs. adaptation, we now consider processes on the group level: split and merge, on the one hand, and influence on the other hand.
To distinguish between them, we introduce a threshold parameter $0\leq \gamma\leq 0.5$
that defines an interval for the group frequency $f_{a}(t)$.
We consider two cases:

\textbf{$f_{a}<\gamma$ or $f_{a}>(1-\gamma)$: }
  This means, there is a clear majority in the group, either of agents with opinion 1 $(f_{a}>1-\gamma)$ or of agents with opinion 0 $(f_{a}<\gamma)$.
In this case the majority can influence the minority such that each agent belonging to the minority changes its opinion toward the majority opinion with a transition rate given by Eqn.~\eqref{eq:7a}.
  Agents with the majority opinion will not change their opinion.
  Specifically:
\begin{align}
  w(1-s_{i}|s_{i}) &\propto  f_{a}^{1-s_{i}} &&  w(s_{i}|1-s_{i}) =0  & \text{if } f_{a}^{1-s_{i}} > f_{a}^{s_{i}} \nonumber \\
   w(1-s_{i}|s_{i}) &=0 && w(s_{i}|1-s_{i}) \propto  f_{a}^{s_{i}} & \text{if } f_{a}^{1-s_{i}} < f_{a}^{s_{i}}
  \label{eq:77a}
\end{align}
In case of a tie, i.e., $f_{a}$=0.5, the ``minority'' opinion is chosen randomly.
Note that this influence not necessarily leads to group consensus because the transitions occur only with a certain probability.

\textbf{$\gamma<f_{a}<(1-\gamma)$:}
  This means, the fractions of agents with opinion 1 and of agents with opinion 0 are of comparable size, i.e., the majority is not sufficiently large and it does not dominate the group.
  In this case the group will \emph{split} into two smaller groups, one for each opinion.
  If the former group has size  $n_{a}$ and fraction $f_{a}$, the two new groups $a_{1}$ and $a_{2}$ have the sizes ${n_{a_{1}}}=f_{a}n_{a}$, ${n_{a_{2}}}=[1-f_{a}]n_{a}$ and the composition $f_{a_{1}}=1$ and $f_{a_{2}}=0$.

  Subsequently, these two groups will \emph{merge} with other groups $b$ and $c$
  that have their opinion as the majority opinion.
  For instance, if group $b$ has size ${n_{b}}$ and $f_{b}>0.5$,
  then group $a_{1}$ will merge with $b$, to obtain a new group $b_{1}$ with
  \begin{align}
    \label{eq:9a}
    {n_{b_{1}}}={n_{b}}+{n_{a_{1}}}\;; \quad
    f_{b_{1}}=\frac{f_{b}{n_{b}}+{n_{a_{1}}}}{{n_{b}}+{n_{a_{1}}}}
  \end{align}
  Group $a_{2}$ however will merge with group $c$ if $f_{c}<0.5$, and we have a new group $c_{1}$ with
\begin{align}
    \label{eq:9b}
    {n_{c_{1}}}={n_{c}}+{n_{a_{2}}}\;; \quad
    f_{c_{1}}=\frac{f_{c}{n_{c}}}{{n_{c}}+{n_{a_{2}}}} 
  \end{align}
  The groups $b$ and $c$ are randomly chosen from those groups that fulfill the conditions $f_{b}>0.5$ and $f_{c}<0.5$.
  The two new groups that result from the merger become hyperedges in the modified hypergraph.

\section{Results of computer simulations}
\label{sec:results-comp-simul}

\subsection{Heterogeneous Mean Field Approximation}
\label{sec:heter-mean-field}

With the adaptive voter model specified for a hypergraph, we are now interested whether the system will reach consensus, characterized by a magnetization $\abs{M}=1$.
We particularly study how consensus depends on the initial configuration of the hypergraph, that is, on the distribution of \emph{group sizes} and on the bias in the initial \emph{opinion distribution}.
In this section we perform numerical simulations, while in the next section we present two analytical studies.

For the initial distribution of opinions, we use the bias $\mean{M(0)}$ and randomly assign opinions such that the expected fraction of agents with opinion 1 is $[1+\mean{M(0)}]/2$.
To assign agents to groups of different sizes $n_{a}$, 
we assume that group sizes vary and follow a Poisson distribution on the support $n\in[2,\infty)$:
\begin{align}
  \label{eq:10}
  \pi(n)=\frac{1}{\beta}e^{-\frac{n-2}{\beta}}
\end{align}
This distribution is chosen because we assume that establishing larger groups should be more costly and therefore happen less frequently. 
The parameter $\beta\in[0,\infty)$ will be varied in the following. 
It determines the mean group size $\mean{\pi(n)}=2+\beta$ and the standard deviation is $\sigma_{\pi(n)}=\beta$. 
That means, the higher the  $\beta$ the more likely it is that large groups are formed. 

To study the dynamics of the hypergraph, we proceed as follows:
At each time step, we sample the group size $n_{a}$ from the exponential distribution.
We then randomly assign $n_{a}$ agents to this group, to form a hyperedge.
To this group, we apply the rules specified in Section~\ref{sec:dynamics} and then move to the next time step. 
We continue until an equilibrium is reached, which means that the system has reached consensus.
This state can indeed be achieved because we do not restrict interactions.

Our sampling procedure varies a standard technique in epidemiology called the heterogeneous mean field approximation \cite{pastor2001epidemic}.
Firstly, we assume that all agents can interact and therefore can potentially be in the same group.
If the number of groups is large, this is akin to a mean field approximation, in which every agent interacts with every other agent. 
Secondly, over time agents have been assigned to different groups of various sizes.
This reflects the heterogeneity.
Over time  groups overlap if they contain agents previously assigned to other groups, i.e. hyperedges overlap and form the hypergraph.
Because of the mean field approximation, we can assume that the distribution of group sizes $\pi(n)$ is not affected by the dynamics on the hypergraph, and that the split-merge process of the dynamics, in case of $\gamma < f_a < (1-\gamma)$, is essentially  absent.

\subsection{Dynamics of magnetization}
\label{sec:evol-magn}

\begin{figure}[htbp]
    \centering
\includegraphics[width=0.4\textwidth]{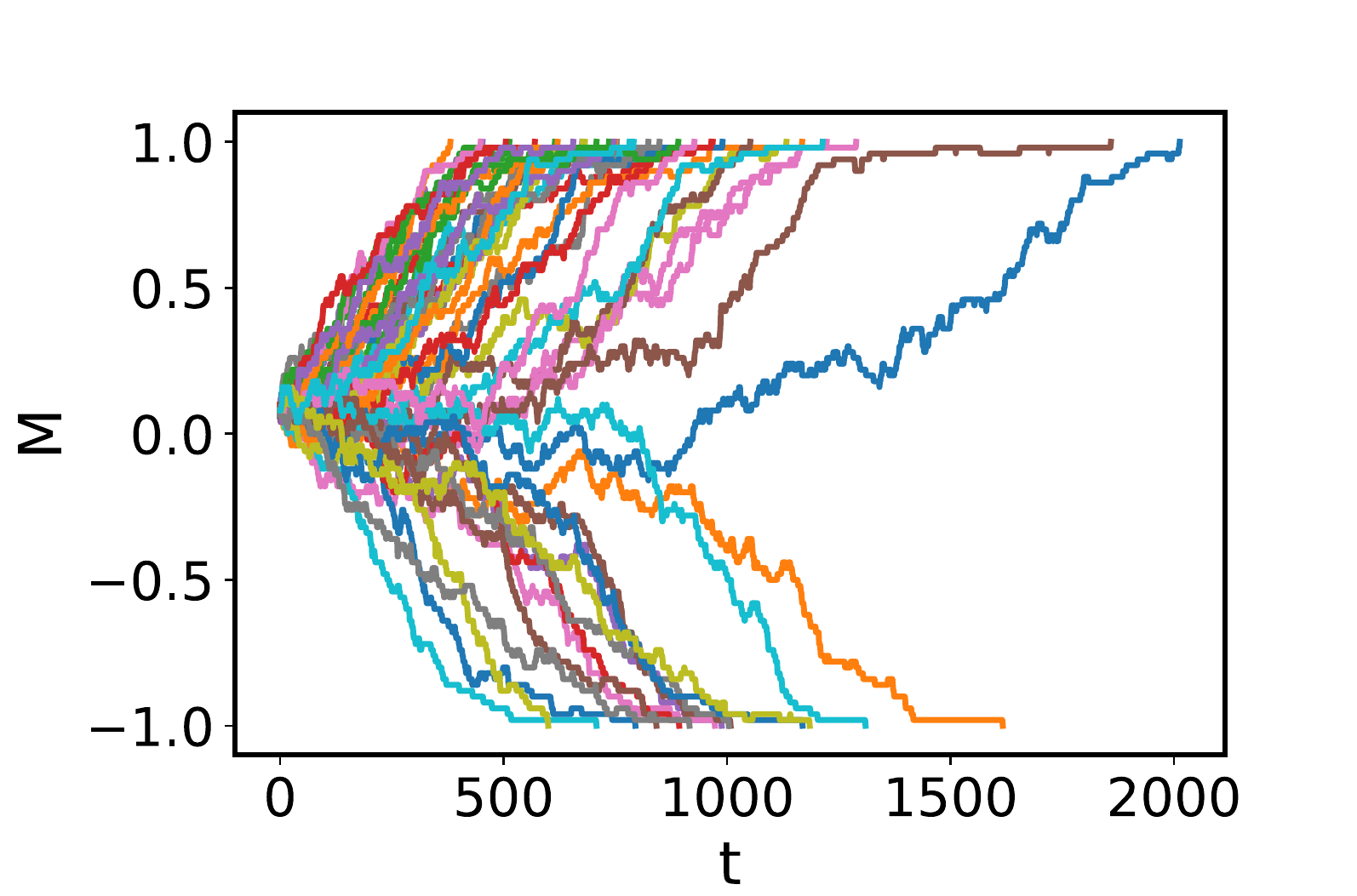}
    \caption{Dynamics of the magnetization for 50 simulations starting from the same initial configuration with $M(0)=0.08\neq 0$. Parameters: $N=100$, $p=0.5$, $\beta=1.5$, $\gamma=0.3$. 
    }
    \label{fig:hmf_trajectories}
\end{figure}

To study the convergence to consensus on the hypergraph we use the magnetization $M(t)$, Eqn.~\eqref{eq:5}.
The time to consensus, $t_{eq}$, is given by the first time step in which
 $\abs{M(t_{eq})}=1$.
Given an initial configuration with a positive bias  we expect the system to quickly reach an equilibrium with $M=1$. 
But fluctuations might impact the system such that instead the equilibrium at $M=-1$ is reached.
Figure~\ref{fig:hmf_trajectories} shows the respective dynamics for a fixed initial configuration of opinions.
We see that the system always reaches total consensus.
Because the initial magnetization $M(0)$ is small, but positive, for the majority of the simulations the final magnetization is $M=+1$, but fluctuations allow also convergence to $M=-1$.

\begin{figure}[htbp]
    \centering
\includegraphics[width=0.4\textwidth]{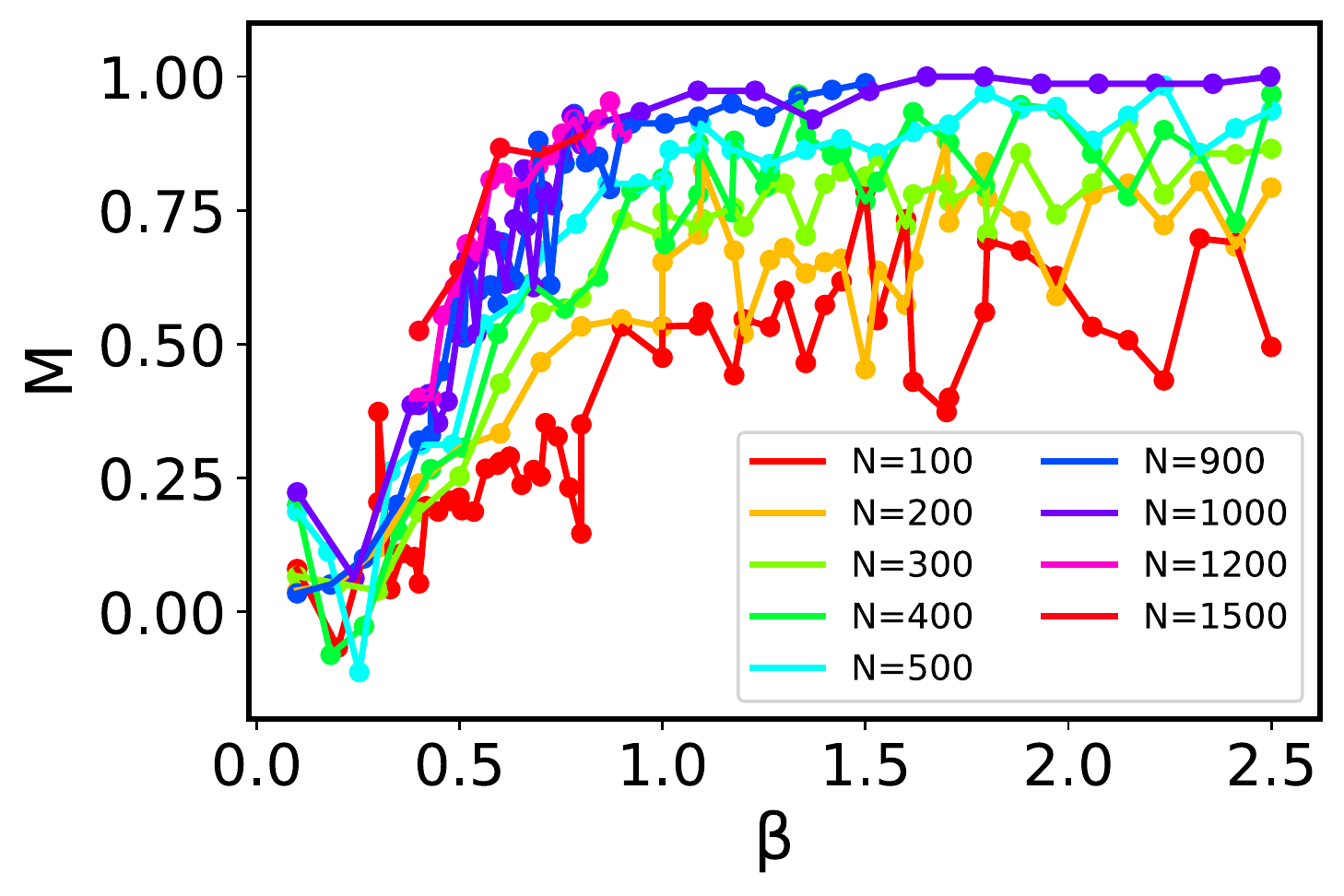}
    \caption{Final average magnetization versus $\beta$ for various values of $N$ for $\mean{M(0)}=0.1$. Each data point is averaged over $20$ trajectories and $10$ initial configurations.}
    \label{fig:FSA_01}
\end{figure}

In Figure~\ref{fig:FSA_01}, we show how the average final magnetization
depends on the heterogeneity of the group sizes, expressed by $\beta$.
Given the initial bias of $\mean{M(0)}=0.1$, we see that for a low group size heterogeneity the final magnetization fluctuates around 0.1.
But with increasing heterogeneity it approaches values close to 1.0.
This has two implications.
First, we see that the initial magnetization is not conserved as in the simple linear voter model.
There, we would expect a final average magnetization equal to the initial one, as we see it for small values of $\beta$. 
However, in our model with increasing $\beta$ we observe a \emph{drift} in the average magnetization away from the initial value, towards +1.
At the same time, Figure~\ref{fig:FSA_01} shows that the average final magnetization does not  always converge to $\mean{M}=+1$, dependent on the system size. 
The observation means that a non negligible fraction of trajectories ends up with the opposite magnetization $M\to (-1)$.
This effect is more pronounced for smaller system sizes and could indicate both finite size effects and the impact of noise.
We need to address this problem in Section~\ref{sec:analytic-results}. 

Secondly, even though the \emph{average} magnetization stays between $-1$ and $+1$, the system in every single run reaches consensus as also shown in Figure~\ref{fig:hmf_trajectories}.
With increasing values of $\beta$, this consensus is more biased towards opinion +1, which was supported by the initial condition  $\mean{M(0)}=0.1$.
Hence, group interactions obviously amplify the small initial bias. 
This is to be expected because with increasing  $\beta$ we sample more agents in each time step.
The average group size becomes larger, which means it is less likely that the minority opinion would be reinforced.
Instead this process reinforces the  majority already present and, hence, increases the final average magnetization.
We have verified that  this drift effect is increased if the initial bias $\mean{M(0)}$ becomes larger.

\begin{figure}[htbp]
    \centering
    (a)\includegraphics[width=0.4\textwidth]{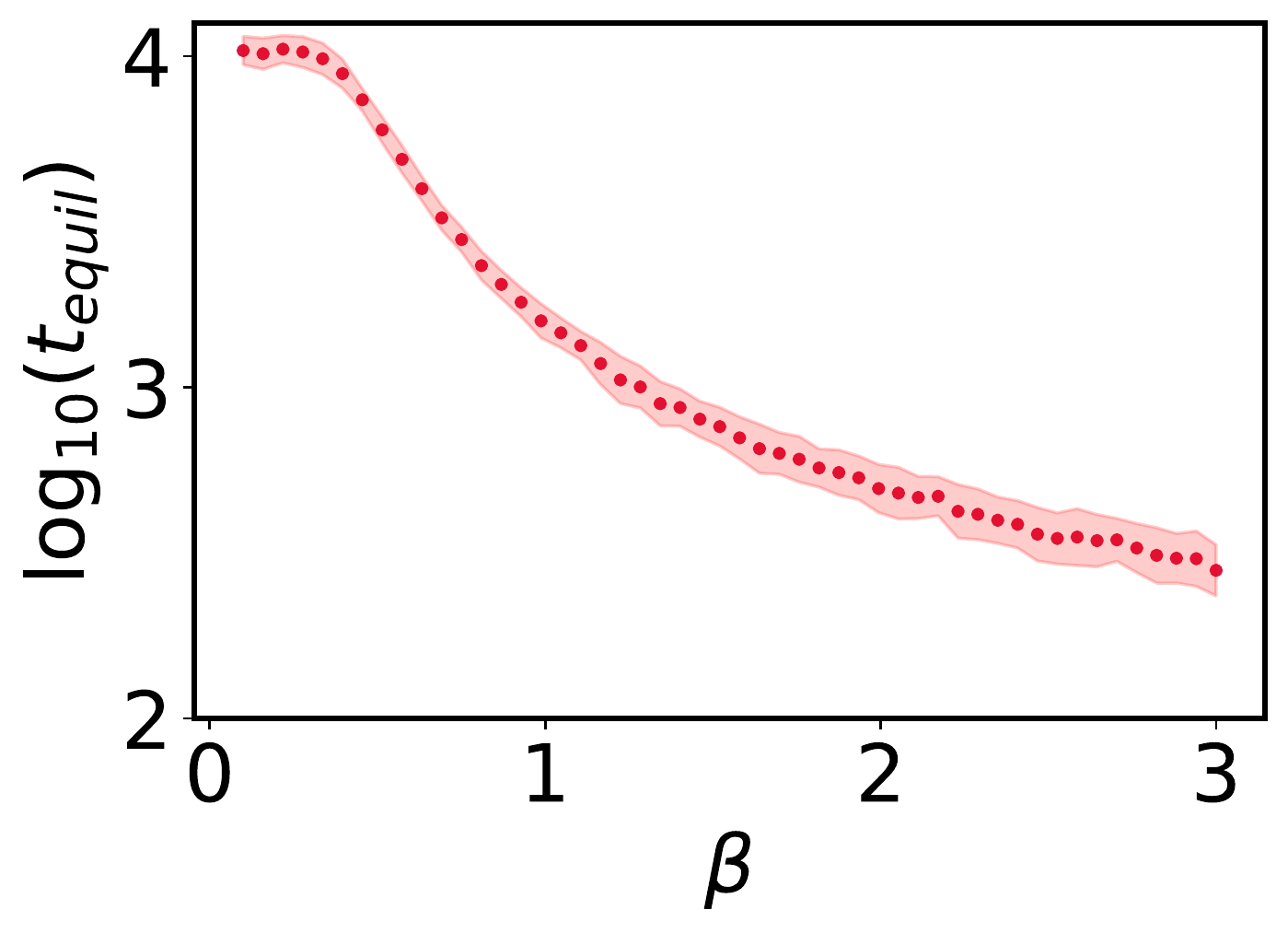} \hfill   (b)\includegraphics[width=0.4\textwidth]{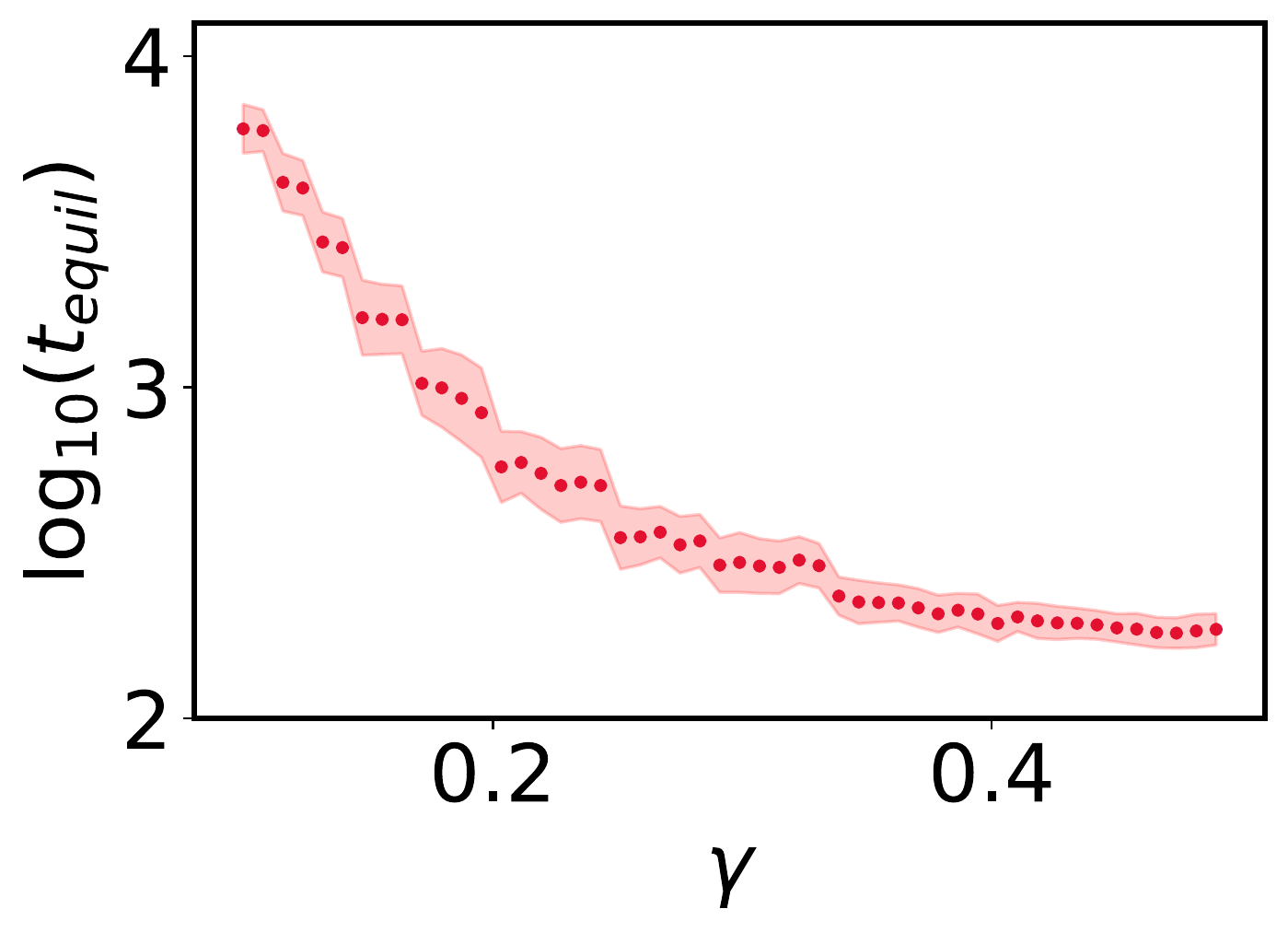}
\caption{Time to equilibrium (log scale) versus (a) $\beta$ and (b) $\gamma$.
  Parameters:  $N=100$,  $r=0.5$, , $\mathbb{E}[m_0]=0.1$, (a) $\gamma=0.3$, (b) $\beta=3$. 
  The points and the shaded area are the average and the standard deviation of the time of convergence for $100$ trajectories of $50$ initial configurations for a given $\gamma$.
}
    \label{fig:t_vs_beta}
\end{figure}

In Figure~\ref{fig:t_vs_beta} we study how quickly the system reaches its equilibrium dependent on the two model parameters $\beta$ and $\gamma$.  
We find that, for fixed $\gamma$, the time to equilibrium $t_{eq}$ decreases with $\beta$, i.e. group interactions accelerate the convergence to consensus.
This is not trivial because during group interaction agents adopt the opinion of the majority \textit{only} if the influence process takes place.
This happens only if $\gamma$ is large enough, i.e. if there is already a strong majority in the group with $f_{a}<\gamma$ or $f_{a}>(1-\gamma)$. 
When $\gamma$ is small, the split-merge process becomes more likely.

To explore the effect of $\gamma$ on the time to reach consensus, we plot $t_{eq}$ versus $\gamma$ in Fig.~\ref{fig:t_vs_beta}(b).
For high values of $\gamma$ the system converges significantly faster because the influence process dominates.
A closer inspection shows a stair like decrease, i.e. there are some small ranges of $\gamma$ in which the $t_{eq}$ is almost constant.
This is due to the discrete size of groups that imply $f_{a}$ to be a rational number.

The computer simulations have motivated to further study the impact of the model parameter $\beta$ instead of $\gamma$.
The latter only indirectly determines group effects on the opinion dynamics by specifying whether influence takes place. 
$\beta$, on the other hand, explicitly affects the opinion dynamics via the heterogeneity of group sizes in a probabilistic manner.
In the following, we present two analytical analyses about the influence of $\beta$.
First, we develop an analytical approximation for the average magnetization using the master Eqn.~\eqref{eq:6}.
This will assume that agents update their opinions independent from another in groups of varying sizes.  
As a second step, we improve this approximation by using the Kramer escape formula.

\section{Analytic results}
\label{sec:analytic-results}
\subsection{Derivation of the expected magnetization}
\label{analysis_sect}

We now develop an analytical approximation for the dynamics of the expected magnetization $\mathbb{E}[M(t)]=\mean{M(t)}$ of the system in the limit of the HMF dynamics used above.
This requires us to average over all individual stochastic dynamics given by the master Eqn~\eqref{eq:6}.
To do this, we assume that at each time step agents update their opinions independently.
Hence, we move our focus away from groups towards agents as the units of analysis, to make the summations required for the averaging procedure tractable.

Let us randomly sample one focal agent out of $N$ agents and a group size $n$ from the distribution $\pi(n)$, Eqn. \eqref{eq:10}.
Agents change their opinions dependent on the group.
Thus, we first have to determine whether the agent belongs to the group of size $n$.
In the HMF limit groups can be composed of possibly all agents.
Thus, we have to consider that there are $\binom{N}{n}$ different groups of size $n$.
The number of groups that include the focal agent is then $\binom{N-1}{n-1}$.
Hence the probability that the focal agents belongs to a randomly chosen group is
$\binom{N-1}{n-1}/\binom{N}{n} = n/N$, which coincides with the probability of a random sampling.

We assume that the focal agent changes its opinion from 0 to 1, which is expressed by the transition rate $w(1|0)$.
This rate is composed of a term $w(0|1,n)$ that describes the opinion change given the presence of the group of size $n$ and a second term that depends on the probability $\pi(n)$ to have a group of size $n$ times the probability that the focal agent is part of it.
In combination we obtain:
\begin{align}
\label{eq:general}
    w(1|0)= \sum_{n=2}^{N} \pi(n) \frac{n}{N}w(1|0, n)
\end{align}
Depending on $n$, the focal agent has either a pairwise interaction ($n=2$) or a group interaction ($n\geq 3$).
For $n=2$, the focal agent can change its opinion only if the other agent has the opposite opinion \emph{and} if an adoption of the new opinion, rather than rewiring, occurs. 
The first condition is met with probability $Nf/(N-1)\approx f$ because in the HFM regime the other agent is chosen uniformly at random among the remaining $N-1$ agents.
The second condition is met with probability $(1-r)$.
In combination we obtain for pairwise interactions:
\begin{align}
  \label{eq:w_pairwise}
  w(1|0, n=2) = f(1-r) 
\end{align}
For $n\geq3$, we have to consider the two different possibilities for group interactions, namely either adoption of the opposite opinion or split-merge processes. 
According to Eqn~\eqref{eq:77a} adoption occurs with a probability $k/n$ where $k$ is the number of agents with opinion 1 in the group, but only if $k/n>1-\gamma$. 
The probability that $k$ agents have opinion 1 in the chosen group of size $n$ is approximated by the binomial distribution $\binom{n-1}{k} f^{k}(1-f)^{(n-1-k)}$.
This holds because the group is formed by randomly grouping $(n-1)$ agents from the remaining $(N-1)$ agents. 
Among these $(N-1)$ agents, a fraction $Nf/ (N-1)\sim f$ has opinion 1 and the remaining fraction has opinion 0.
Hence, $k$ agents have  opinion 1 with probability $f^{k}$ and the remaining $(n-1-k)$ agents with opinion 0 with probability $(1-f)^{(n-1-k)}$.

Finally, the binomial coefficient considers the different possible combinations of having $k$ agents with opinion 1 and $(n-1-k)$ with opinion 0.
Note that, as the focal agent with opinion 0 already belongs to the group, the binomial coefficient allows to choose only up to $n-1$, and not up to $n$, agents with opinion 1. 
In combination the transition rate for the focal agent to adopt the opposite opinion because of group influence becomes:
\begin{align}
  \label{eq:w_group}
    w(1|0, n\geq3) =  \sum_{k=\ceil{n(1-\gamma)}}^{n-1} \binom{n-1}{k} f^{k}(1-f)^{(n-1-k)} \frac{k}{n}
\end{align}
where the summation starts from $k=\ceil{n(1-\gamma)}$ as influence occurs only when $k/n>1-\gamma$.

By insertion of Eqs.~\eqref{eq:w_pairwise},~\eqref{eq:w_group} into Eqn.~\eqref{eq:general}, we finally obtain the transition rate:
\begin{align}
\label{eq:w_0_to_1_final}
  \begin{split}
    &    w(1|0)= \pi(2) \frac{2}{N}(1-r) f \\
    & + \pi(n) \frac{n}{N} \sum_{k=\ceil{n(1-\gamma)}}^{n-1} \binom{n-1}{k} f^{k}(1-f)^{(n-1-k)} \frac{k}{n}
  \end{split}
\end{align}
A similar expression can be derived for the opposite transition rate:
\begin{align}
\label{eq:w_1_to_0_final}
  \begin{split}
    &  w(0|1)= \pi(2) \frac{2}{N}(1-r) (1-f)  \\ &+ \pi(n) \frac{n}{N} \sum_{k=\ceil{n(1-\gamma)}}^{n-1} \binom{n-1}{k} (1-f)^{k}f^{(n-1-k)} \frac{k}{n}
  \end{split}
\end{align}
With these transition rates, we obtain from the master equation in the HMF limit:
\begin{align}
  \begin{split}
    &  \frac{df(t)}{dt} = \frac{dp(1,t)}{dt} = w(1|0) (1-f) -  w(0|1) f  \\ 
    & = f(1-f) \sum_{n=3}^{N}\pi(n) \sum_{k=\ceil{n(1-\gamma)}}^{n-1}
    \frac{k}{N} \times \\ &\times \binom{n-1}{k} \Big[f^k(1-f)^{n-k} - (1-f)^{k}f^{n-k}
    \Big]
  \end{split}
      \label{eq:master_eq_final}
\end{align}
We note that this dynamics has become independent of the rewiring rate $r$.
With Eqn.~\eqref{eq:5} we find for the expected change in the average magnetization:
\begin{equation}
  \begin{split}
&    \mean{M(t+1)-M(t)}=\\
&    2 \Bigg\{\sum_{n=3}^N \pi(n)\frac{1}{N}\sum_{k=\ceil{n(1-\gamma)}}^{n-1}k\binom{n-1}{k} \times\\
    \times \Bigg[&\bigg(\frac{1+\mean{M(t)}}{2} \bigg)^{k}\bigg(\frac{1-\mean{M(t)}}{2}\bigg)^{n-k}-\\
    & \bigg(\frac{1-\mean{M(t)}}{2}\bigg)^{k}\bigg(\frac{1+\mean{M(t)}}{2}\bigg)^{n-k}\Bigg]\Bigg\}
      \end{split}
  \label{eq:magnet_analyt}
\end{equation}
A further simplification of this equation is given in the Appendix.
From Eqn.~\eqref{eq:magnet_analyt}, it can be easily shown that for $0<\mean{M(t)}<1$ the summand is positive and for $-1<\mean{M(t)}<0$ negative, respectively, for every value of $k$ since $\gamma\leq \frac{1}{2}$.
This means that as long as there are higher order interactions ($\beta\neq0$) the only fixed points of the dynamics are $\mean{M}=0$, $\mean{M}=1$ and $\mean{M}=-1$.
Based on the signs, $\mean{M}=0$ is unstable, while $\mean{M}=\pm1$ are stable fixed points.
Therefore, starting with $\mean{M(0)}>0$, we should expect that the magnetization averaged over many trajectories and initial configurations always goes to total consensus with opinion $1$ for $\beta\neq0$. 

These expectations can now be compared with the results of computer simulations for various system sizes already shown in Figure~\ref{fig:FSA_01}.
There we found that the average final magnetization does not converge to $\mean{M}=\pm 1$, but stays between $-1$ and $+1$.
The effect is more pronounced for small values of $\beta$, but remains even for high values of $\beta$, dependent on the system size.

This leaves us with two hypotheses about the reasons for the deviations from the analytical expectations: (i) existence of a finite-size phase transition with $\beta$ as the order parameter, (ii) influence of noise.
To check hypothesis (i), we followed the method explained in \cite{privman1990finite}.
We applied the finite size analysis to the data depicted in Figure~\ref{fig:FSA_01} and to additional data for a different $\mean{M(0)}=0.35$.
We verified that there is no combination of critical exponents and  critical values of the order parameter that would result in a scaling.
This implies that there is no finite-size phase transition with $\beta$ as the order parameter.  
 
Prompted by this, we have to investigate hypothesis (ii), whether the curves of the plots in Figure~\ref{fig:FSA_01} can be explained by noise from sampling finite systems.

\subsection{Analysis with noise}
\label{analysis_with_noise_sect}

The magnetization of the initial configurations follows a binomial distribution with mean $\mean{M(0)}$ and non-zero variance. Therefore, some initial configurations are more prone to ``switching'' their magnetization, while others are more robust.
To estimate this noise effect, we calculate the probability that a  magnetization trajectory changes its sign starting from an average initial magnetization $\mean{M(0)}$.

We consider a stochastic force on the magnetization trajectory.
Assuming that each trajectory is independent, this translates to adding a stochastic force to the magnetization dynamics averaged over many initial configurations and trajectories, as given by Eqn.~\eqref{eq:magnet_analyt}. 
We study the dynamics in the continuum limit $\mean{M(t+1)-M(t)}\approx d\mean{M(t)}/{dt}$.
This is a reasonable approximation since the time to converge to consensus is of the order $\mathcal{O}(10^3)$, as shown in Figure~\ref{fig:t_vs_beta}.
The RHS of Eqn.~\eqref{eq:magnet_analyt} is a one-dimensional function of $\mean{M(t)}$ and can therefore be expressed as a conservative force resulting from the gradient of a potential $U(\mean{M})$.
Eqn.~\eqref{eq:magnet_analyt} can then be rewritten as an overdamped Langevin equation:
\begin{equation}
    \lambda \frac{d\mean{M(t)}}{dt}= -\frac{dU(\mean{M})}{d\mean{M}}+\eta(t)
    \label{dmdt_noise},
\end{equation}
where $\lambda$ is the damping coefficient. $\eta(t)$ is Gaussian noise with 
$\langle \eta(t) \eta(t') \rangle = 2D\delta(t-t')$
where $D={k_{B}T}/{\lambda}$ is the diffusion constant and $k_{B}T$ is the effective temperature.

To make $U(\mean{M})$ independent of finite size effects, we use the fact that the series in Eqn.~\eqref{eq:magnet_analyt} converges for $n\to \infty$.
We further assume $N\rightarrow \infty$, this way the factor ${1}/{N}$ in Eqn.~\eqref{eq:magnet_analyt} is absorbed into the damping coefficient $\lambda$ and  the effective temperature $k_{B}T$.
$dU(\mean{M})/{d\mean{M}}$ is then given by the RHS of Eqn.~\eqref{eq:magnet_analyt} without the term $1/N$. 
The values of $\lambda$ and $k_{B}T$ are not known, however they are expected to depend only on the quantities $N$ and $\beta$ given that the other parameters stay constant; i.e. $\lambda=\lambda(N,\beta)$ and $k_{B}T=k_{B}T(N, \beta)$.

From the discussion of Eqn.~\eqref{eq:magnet_analyt}, we know that the dynamics has three fixed points, two stable ones at $\mean{M}=\pm1$ and one unstable at $\mean{M}=0$.
Therefore, the potential $U(\mean{M})$ has a double-well shape with the potential barrier at $\mean{M}=0$.
Our initial question whether a magnetization trajectory could change its sign can now be recast as an escape from a potential well.
This is a  classical problem in statistical physics, described by 
Kramer's escape rate formula \cite{risken1996fokker}.
It allows to estimate the escape probability by means of an activation energy $E_b \gg k_{B}T$ and the second derivative of the potential $U$ taken close to one of the potential minima. 

At difference with Kramer's set-up, in our case the trajectories at $t=0$  are not at near the bottom of a well, but close to the top of the barrier.
However, based on the derivation for the Kramer's escape rate formula \cite{risken1996fokker} it can be shown that the formula is still applicable for $E_b \gg k_{B}T$ and $\mean{M}$ close enough to the top of the barrier.
Because in our case $\mean{M(0)}=\mean{M}=0$, we further approximate the curvature of the potential as $U''(\mean{M(0)})\approx U''(0)$.
With this, Kramer's escape rate, which is in our case the rate $S$ of switching the magnetization, becomes:
\begin{equation}
    \begin{split}
S=\frac{|U''(0)|}{2\pi\lambda(N,\beta)} e^{-\frac{E_b}{k_{B}T(N,\beta)}}
    \end{split}
    \label{kramer_our_case}
\end{equation}
where $E_b=U(0)-U(\mean{M(0)})$, i.e. $E_{b}$ is measured in terms of the initial magnetization $\mean{M(0)}$.
We calculated the curvature analytically by differentiating $U(\mean{M})$ with respect to $\mean{M}$, while the energy barrier $U(0)$ was obtained from numerical integration.

\begin{figure}
    \centering
\includegraphics[width=0.4\textwidth]{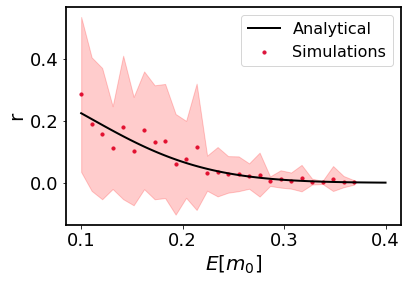}
\caption{Switching rate $S$ versus initial magnetization $\mean{M(0)}$.
  Parameters: $N=100$, $r=0.5$, $\beta=1.5$, $\gamma=0.3$. The red points and the shaded area are the mean and the standard deviation of the switching rate averaged over $50$ trajectories and $30$ initial configurations. The black line is the best fit ($R^{2}=0.88$) of $S$, Eqn.~\eqref{kramer_our_case}, with
  $k_{B}T=0.010\pm0.007$ and $\lambda=0.45\pm0.02.$
  }
    \label{kramer_fit}
\end{figure}

To relate the analytic expression for the switching rate, Eqn.~\eqref{kramer_our_case}, to our computer simulations, we chose 
an initial positive magnetization $\mean{M(0)}>0$ and simulated many trajectories for varying initial configurations with fixed $N$ and $\beta$. 
The switching rate is then given by the fraction of those trajectories whose final magnetization has become negative.
An example is shown in Figure~\ref{kramer_fit} together with the best fit for Eqn.~\eqref{kramer_our_case}.

\section{Numerical results}
\label{sec:numerical-comparison}

\subsection{Scaling relations for $\lambda$ and $k_{B}T$}
\label{sec:scal-relat-lambda}

For a complete understanding of the system dynamics we still have to determine the scaling relations of $k_{B}T(N,\beta)$ and $\lambda(N,\beta)$ dependent on the system size $N$ and the group heterogeneity $\beta$.
To obtain these relations, we used parallel computing facilities and repeated the procedure behind Figure~\ref{kramer_fit} for varying values of $\beta\in[1.0,3.0]$ and for $7$ different values of $N$.
For every $\beta$ and $N$, the fitted values of $kT$ and $\lambda$ were used to test that the two assumptions to apply Kramer's escape rate formula were satisfied, i.e. $U''(\mean{M(0)})\approx U''(0)$ and $E_{b}\gg k_{B}T$.

\begin{figure}[htbp]
    \centering
(a)\includegraphics[width=0.4\textwidth]{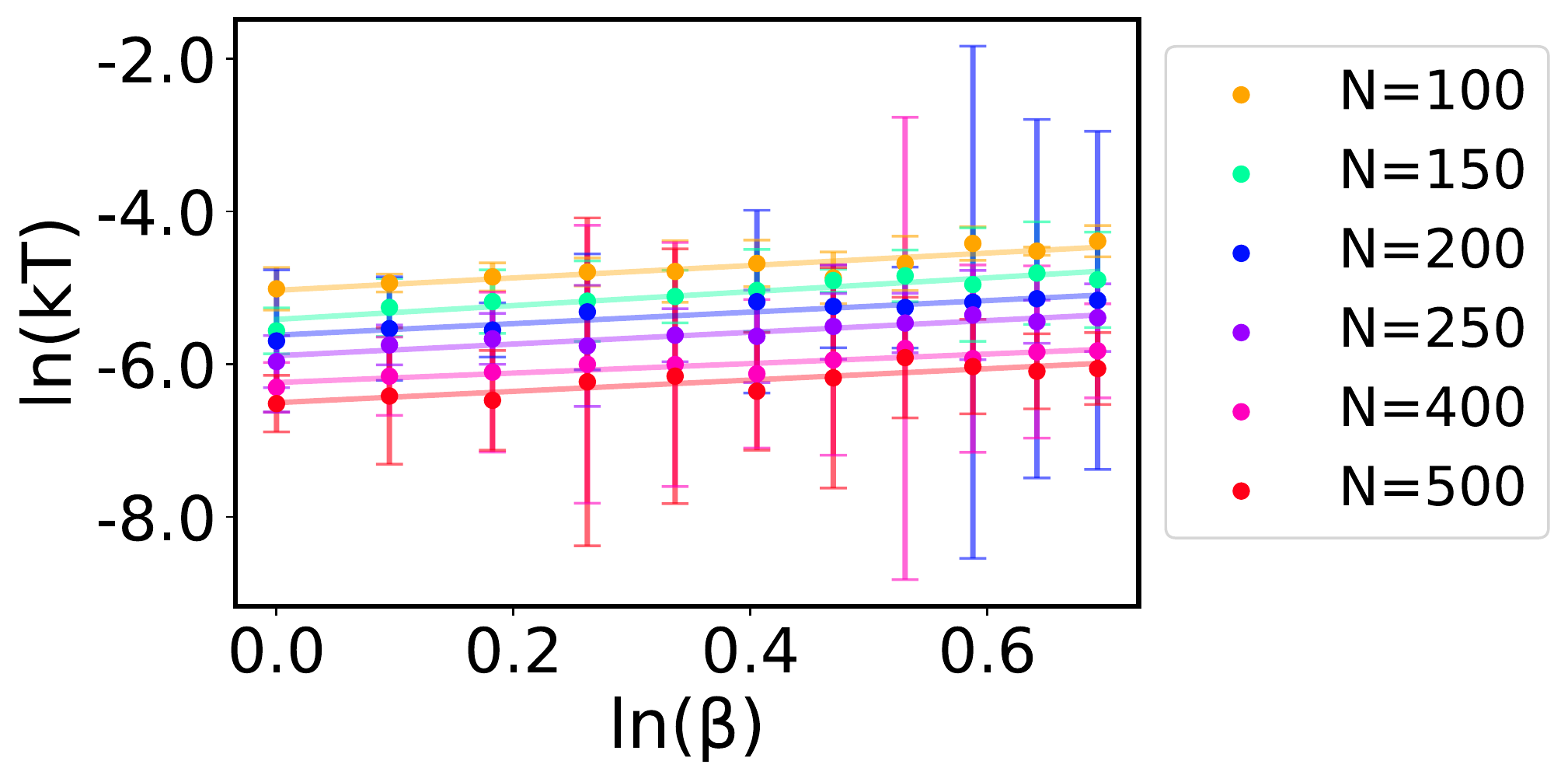}\hfill
(b)\includegraphics[width=0.4\textwidth]{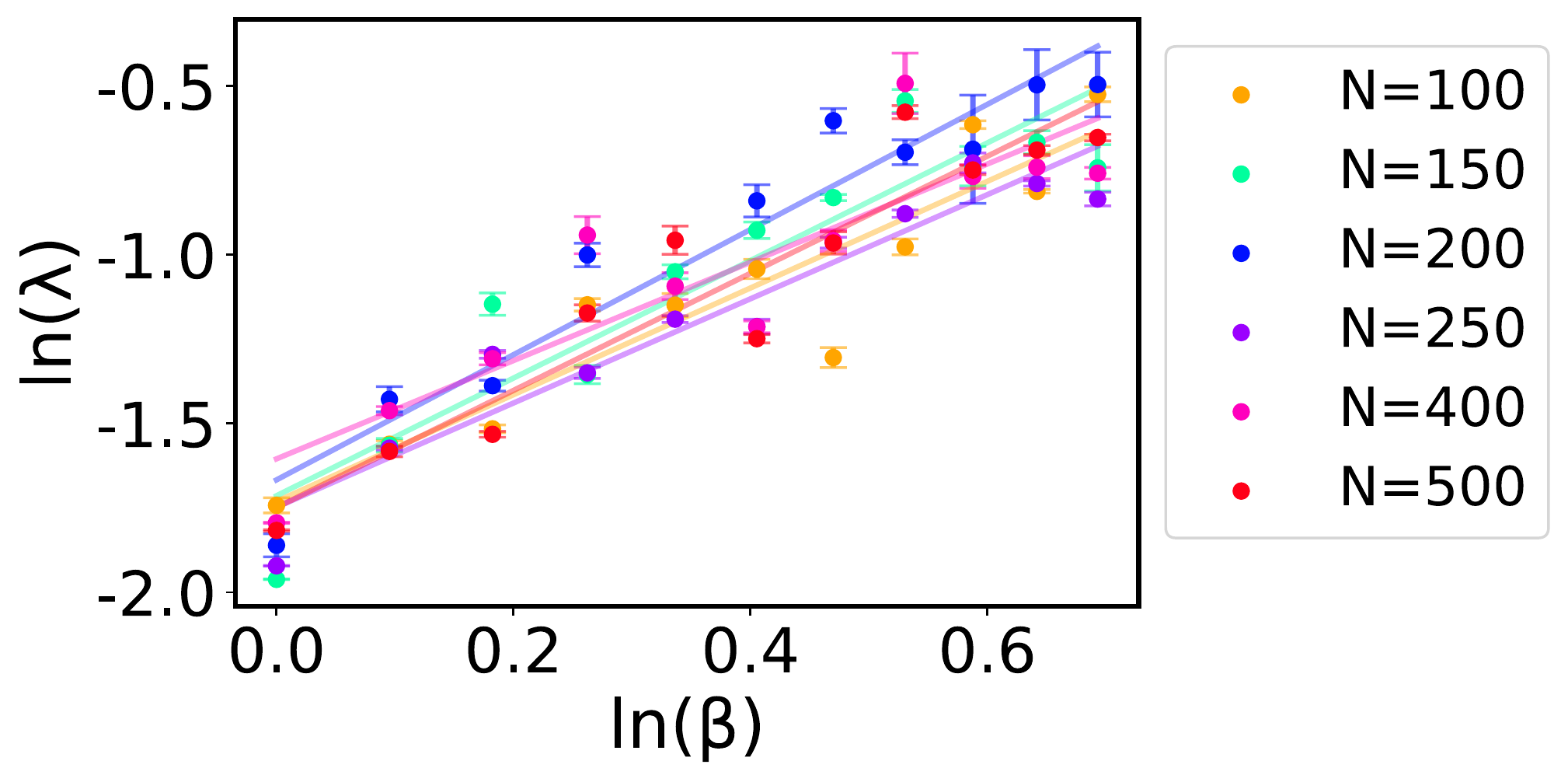}
\caption{ (a) $\ln(kT)$ and (b) $\ln(\lambda)$ versus  $\ln(\beta)$  for $7$ values of $N$ for $r=0.5$ and $\gamma=0.3$. %
  The $R^2$ of the fitting lines are on average 0.85 and at least 0.78 (a) and 0.75 (b). }
    \label{fig:lnkt_vs_lnbeta}
\end{figure}

The results are shown in Figure~\ref{fig:lnkt_vs_lnbeta}.
We verified that the slopes are approximately equal for every value of $N$ while the intercepts depend on $N$.
Following a similar procedure for constant $\beta$, we found that the slope of $\ln(\lambda)$ and $\ln(kT)$ versus $\ln(N)$ can be accurately approximated to be independent of $\beta$.
Based on these observations, we propose the following scaling relations:
\begin{align}
    k_{B}T(N,\beta) & \propto  e^\delta N^\nu \beta^{\mu}, \nonumber \\
    \lambda(N,\beta)& \propto  e^\zeta N^\theta \beta^{\kappa}
  \label{kT}
\end{align}

For the rewiring probability $r=0.5$ and $\gamma=0.3$, the values of the coefficients were calculated as:
\begin{align}
  \mu=& +0.768\pm0.042\;\;& \kappa=& +1.651\pm0.060 \nonumber \\
  \nu=& -0.882\pm0.001\;\;&  \theta=& -0.004\pm0.085 \nonumber \\
  \delta=& -0.983\pm0.002\;\;&  \zeta=& -1.698\pm0.136 \nonumber
  \label{eq:12}
\end{align}
These scaling relations can now be used to calculate $S$, Eqn.~\eqref{kramer_our_case}, for a large range of $N$ and $\beta$.
It is important to note that the scaling relations indeed only hold as long as the above two assumptions to apply Kramer's escape rate formula were satisfied.
We tested that the power law scaling breaks down in parameter regions where the two assumptions do not hold.

\subsection{Impact of the switching rate}
\label{sec:impact-switch-rate}

\begin{figure}[htbp]
    \centering
(a)\includegraphics[width=0.4\textwidth]{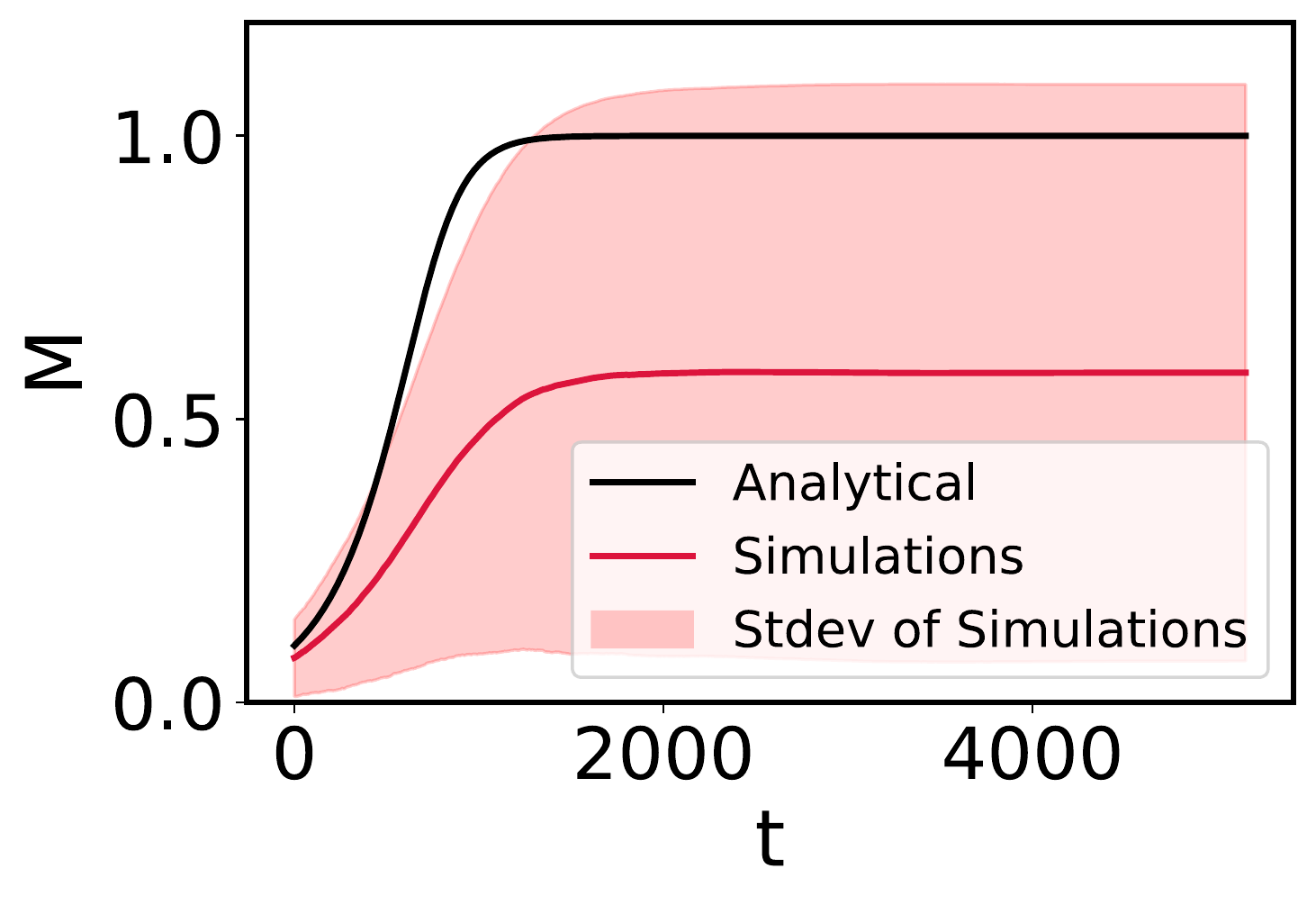}\hfill
(b)\includegraphics[width=0.4\textwidth]{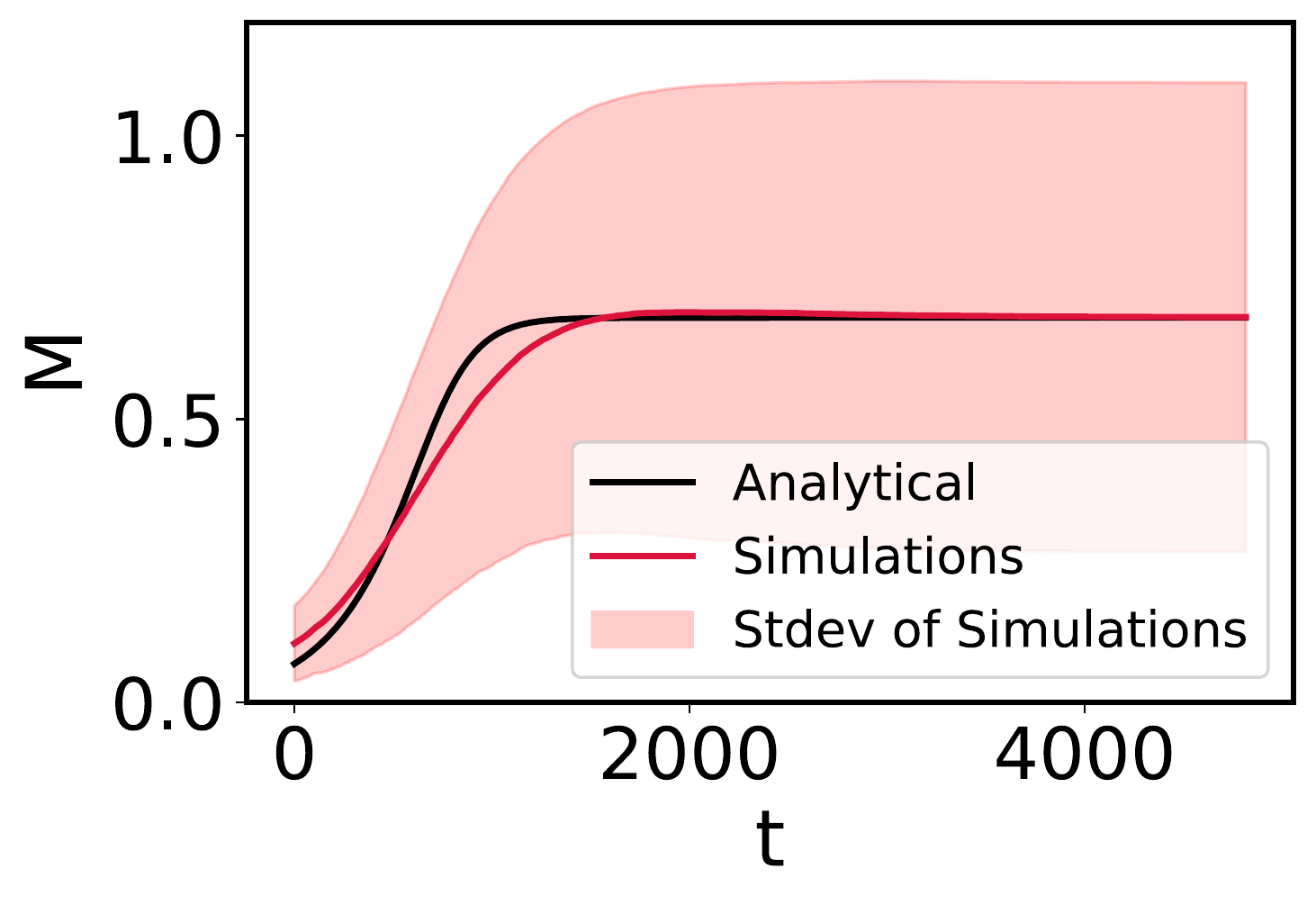}
\caption{Average magnetization, Eqn.~\eqref{mixed_m} (black) and from simulations (red) over time. (a) $S=0$, (b) $S$ from Eqn.~\eqref{kramer_our_case}.
Parameters: $N=200$, $\mean{M(0)}=0.1$, $\beta=1.5$, $\gamma=0.3$, $r=0.55$. 
We run $100$ simulations starting from $50$ different initial configurations. 
}
    \label{fig:evol_noise}
\end{figure}

Now that we know the switching rate $S$, Eqs.~\eqref{kramer_our_case}, \eqref{kT}, of a magnetization trajectory due to noise, we can calculate its impact on the average magnetization, $\mean{M(t)}$, given by Eqn.~\eqref{eq:magnet_analyt}.
The corrected average magnetization is:
\begin{equation}
    \mean{{M}^{c}(t)}=[1-2S] \mean{M(t)}
    \label{mixed_m}
\end{equation}
The factor $[1-2S]$ comes from the fact that the fraction of trajectories that change their sign due to noise increases negative magnetization ($S$), but also decreases the fraction of trajectories that contribute to positive average magnetization ($[1-S]$).
In Figure~\ref{fig:evol_noise} we plot the analytical result for the dynamics of the average magnetization, Eqn.~\eqref{mixed_m} without noise ($S=0$) and with noise ($S\neq0$) and compare it to the dynamics obtained from computer simulations.
Obviously, if the impact of noise is considered the analytical model accurately captures the simulations.

The large error bars in Figure~\ref{fig:evol_noise} reflect that  each magnetization trajectory reaches either +1 or -1 at equilibrium.
It can be shown that the standard deviation for the average magnetization at equilibrium, $\mean{M}$, is given by:
\begin{equation}
  \sigma=\sqrt{\Bigg(\frac{1-\mean{M}}{2}\Bigg)\Bigg(\frac{3+\mean{M}}{2}\Bigg)}
  \label{eq:11}
\end{equation}
The analytical model predicts a final average magnetization $\mean{M^{c}}=0.679$, hence the standard deviation resulting from Eqn.~\eqref{eq:11} is $\sigma=0.54$.
Comparing this with 
the standard deviation of the simulations at equilibrium, $\sigma=0.41$, we find a reasonable $75\%$ accuracy.
That means, the large standard deviation shown in Figure~\ref{fig:evol_noise}(b) is an expected effect of the binary values of the final magnetizations and not of the sampling process.

\begin{figure}[htbp]
    \centering
(a)\includegraphics[width=0.4\textwidth]{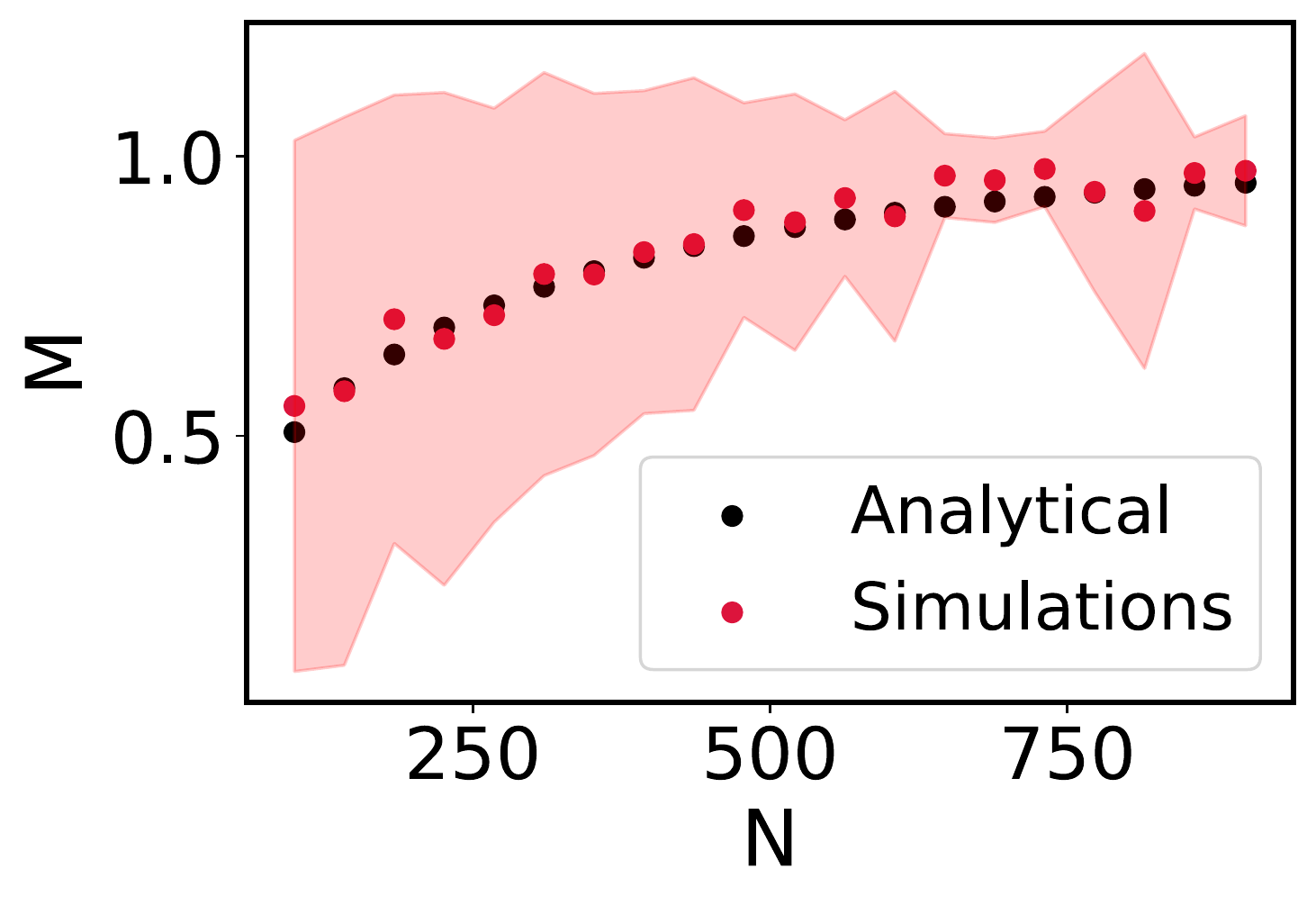}\hfill
(b)\includegraphics[width=0.4\textwidth]{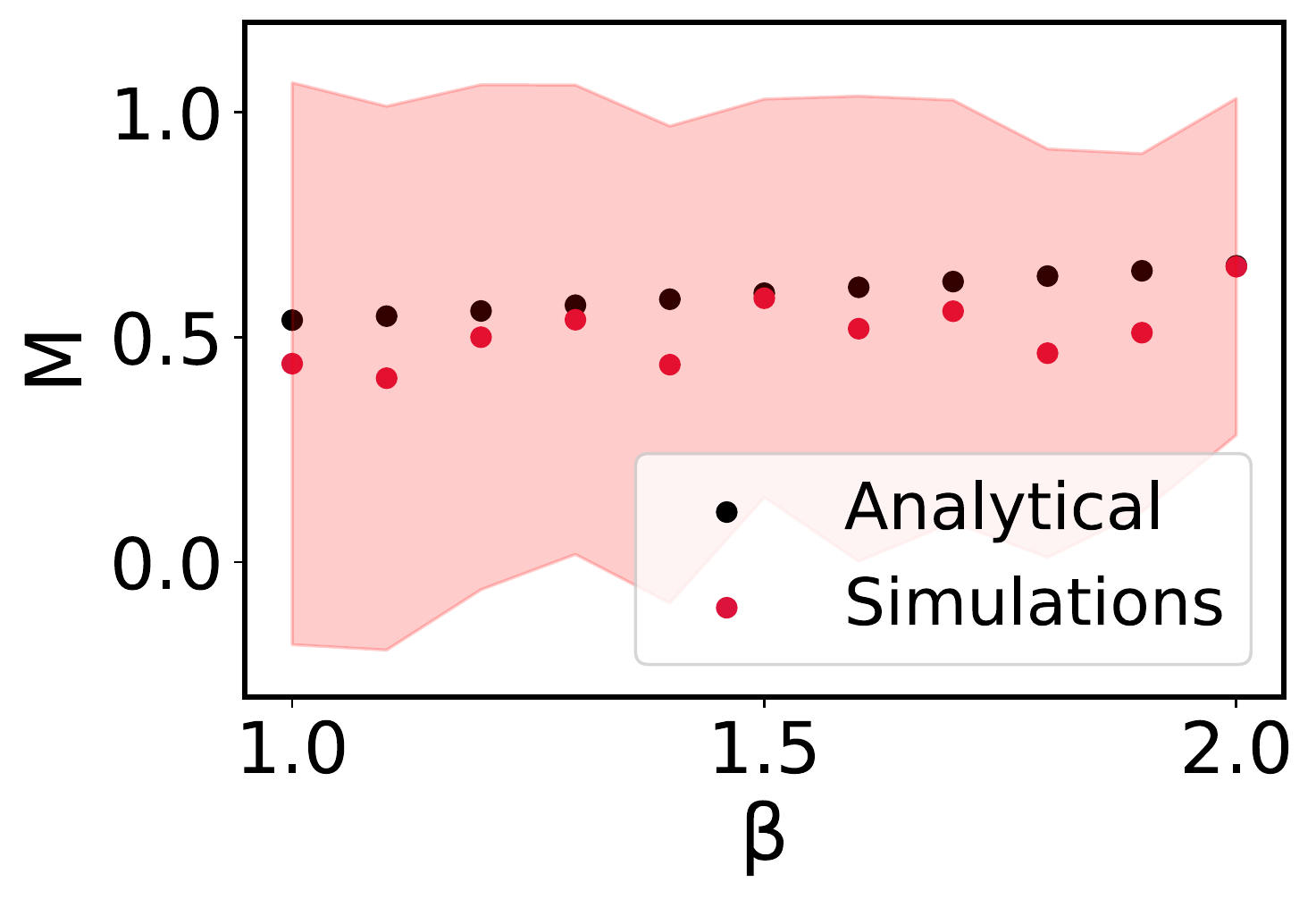}
\caption{Final average magnetization, Eqn.~\eqref{mixed_m} for $t\rightarrow \infty$ (black dots), versus (a) $N$ and (b) $\beta$. Parameters: $\mean{M(0)}=0.1$, $\gamma=0.3$, $r=0.55$, (a) $\beta=1.5$, (b) $N=100$. 
  Red dots and error bars from $100$ simulations starting from $50$ different initial configurations.
  }
    \label{fig:analytical_m_vs_N}
\end{figure}
Eventually, we can also compare the analytical predictions for the final average magnetization with simulations for various $N$ and $\beta$.
The results are shown in Figure~\ref{fig:analytical_m_vs_N}.
The reason for the large standard deviations was already explained above.
But regarding the mean values, the accuracy of the analytical model with respect to the simulations is at a remarkable $83\%$ for Figure~\ref{fig:analytical_m_vs_N}(a) and $75\%$ for Figure~\ref{fig:analytical_m_vs_N}(b).

\section{Conclusion}
\label{sec:conclusion}

Network science has become an essential component to study the dynamics of interacting agents.
Despite its many successes and its flexibility, the research community has begun questioning some underlying assumptions made when representing complex systems by networks, such as the assumption that interactions arise from a combination of pairwise interactions.

In this paper, we have investigated this specific point and aimed at clarifying the role of group interactions in spin systems.
To do so, we have developed an adaptive voter model on hypergraphs.
A hypergraph is composed of hyperedges that represent groups of agents.
These groups can be of different size $n$.
The simplest case $n=2$ covers pairwise interactions.
The voter model assumes that agents are characterized by a discrete spin variable, or ``opinion'', $s_{i}\in\{0,1\}$.
In a network, an agent can change this opinion based on interactions with those agents it shares a link.  
The \emph{adaptive} voter model allows, in addition to the change of opinions, that an agent with a certain probability $r$ rewires its link to another agent instead of adopting the opposite opinion.

If we want to extend this model to a hypergraph, we need to specify (i) how groups of agents influence the individual opinion dynamics and (ii) which processes should replace the rewiring mechanism.
We have proposed that (i) in groups with a clear majority the change of individual opinions depends on the frequency of opinions present in the group, and that (ii) groups without a clear majority will split into smaller groups with a single opinion, which subsequently merge with other groups that support their opinion.

We are interested whether the dynamics on the hypergraph will still allow for consensus, i.e. the domination of only one opinion and, in case, whether this consensus is reached faster or slower compared to the adaptive voter model on a simple network, without group interactions.
Our systemic variable is the magnetization $M(t)$, which reaches +1 if consensus with opinion +1 obtained and -1 if consensus with opinion -1 is obtained.
Values between -1 and +1 indicate the coexistence of opinions.

With our extensions we have introduced new parameters which influence the collective dynamics of reaching consensus and shall be systematically studied. 
In this paper we specifically focused on the parameter $\beta$ that determines the  heterogeneity of group sizes, $n$.
Larger values of $\beta$ allow for larger groups which may more easily reach consensus. 
Additionally, a parameter $\gamma$ controls the re-organization of the hypergraph by determining the threshold for split-merge processes.

Instead of analyzing single computer simulations, we study the \emph{expected} dynamics of the adaptive voter model on hypergraphs.
For the simulations, this requires to average (i) over a large number of initial configurations of opinions and (ii) over a large number of runs.
We also have to address the role of system size, expressed by the total number of agents, $N$, which is varied in the simulations. 
For the analytic investigations, on the other hand, we need to derive a dynamics for the \emph{average} magnetization, starting from the 
initial average magnetization $\mean{M(0)}$. 
The analytic investigations also have to reflect the existence of ``noise'', i.e. randomness in the initial configurations of opinions and deviations from the initial average magnetization.
These problems are solved in our paper using a variant of the heterogeneous mean-field (HMF) approximation \cite{pastor2001epidemic}.
It assumes that groups of varying sizes can be composed from any set of agents, i.e. interactions between agents are not restricted.

Our main findings can be summarized as follows.
First of all, we demonstrate that group interactions accelerate the convergence of the system to total consensus.
As a general insight, this is in agreement with recent results in an adaptive voter model on 2-simplices~\cite{horstmeyer2020adaptive} and in the bounded confidence model on static hypergraphs~\cite{hickok2021bounded}.
We specifically find that small biases in the initial configuration of opinions are more amplified if larger groups can form.
Figure~\ref{fig:FSA_01} shows for small $\beta$ the expected behavior of the normal voter model, but for larger $\beta$ and larger system sizes 
a more pronounced trend toward the biased opinion.
Further, the time to reach consensus drastically decreases with increasing $\beta$ as shown in Figure~\ref{fig:t_vs_beta}(a).
On the other hand, the possibility of split-merge processes in groups, indicated by smaller values of $\gamma$, slows down  the formation of consensus as shown in Figure~\ref{fig:t_vs_beta}(b).
This is reasonable because for smaller values of $\gamma$ imply that the group does not update their opinions.

Secondly, the influence of noise on the dynamics of the average magnetization is quantified such that we could obtain a match between the results of computer simulations and of analytic investigations, as demonstrated in Figure~\ref{fig:analytical_m_vs_N}(b). 
This was made possible by different methodological contributions.
We were able to derive the formal system dynamics for the average magnetization, Eqn.~(\ref{eq:magnet_analyt}), starting from our assumptions for the adaptive voter model on hypergraphs.
These assumptions could, for the HMF limit, be formalized in the transition rates, Eqs.~(\ref{eq:w_0_to_1_final}), (\ref{eq:w_1_to_0_final}), of a master Eqn.~(\ref{eq:6}).
We then corrected in Eqn.~(\ref{mixed_m}) the analytic result for the average magnetization  by deriving a formal expression for the rate at which magnetization trajectories can switch their sign in the presence of noise.
We could provide a scaling relation, Eqn.~(\ref{kT}), for the relevant parameters of the switching rate $S$, Eqn.~(\ref{kramer_our_case}).
This allowed us to calculate the corrections for the average magnetization over a large range of the system parameter $\beta$, as shown in Figure~\ref{fig:evol_noise}(a,b). 
Combined, this lead to the good agreement between simulations and analytics. 
We note that noise only affects the average  magnetization because, in the HMF limit, the magnetization trajectories always go to total consensus.
In general, one should consider that noise also affects the transient phase, where phenomena like slowing down could be expected. 

Possible extensions for future work shall consider heterogeneous initial configurations without applying the heterogeneous mean-field approximation. In this case, consensus would not be the only expected final state, and a phase transition associated to the fragmentation of the system could  emerge~\cite{durrett2012graph}.
We conjecture that the phase transition between total consensus and fragmentation of opinions may be present only for initial configurations with low number of group interactions and therefore low drift of the average magnetization.

Moreover, by using the heterogeneous mean field approximation, we found  that the effect of the probability of rewiring, $r$, is negligible in the dynamics of the average magnetization, Eqn.~(\ref{eq:magnet_analyt}). 
In general, one should expect that the value of $r$ also changes the values of $\lambda$ and $k_{B}T$, which then become functions of $(N,\beta,r)$.
Hence the switching rate $S$ for the magnetization trajectories will change. 

Finally, it would be interesting to investigate the similarities and differences of our model with dynamical systems where the opinions are continuous variables, and not discrete one, for instance by generalizing bounded-confidence model to the case of adaptive hypergraphs.

\small \setlength{\bibsep}{1pt}

\bibliography{arxiv}%

\begin{appendix}
  
\section*{Appendix}
\label{sec:comp-analyt-simul}

For analytical investigations it is useful to simplify the analytic form of the expected magnetization,  Eqn.~\eqref{eq:magnet_analyt}. 
Using the properties for binomial coefficients $k\binom{n}{k}=n\binom{n-1}{k-1}$ and $\sum_{i=0}^n\binom{n}{i}=2^n$ where $n,k\in\mathbb{Z}$, and the abbreviations $\hat{k}=\ceil{n(1-\gamma)}-2$, $\hat{n}=(n-2)$, $\hat{p}_{+}={[1+M(t)]}/{2}$, $\hat{p}_{-}={[1-M(t)]}/{2}$ we obtain  Eqn.~\eqref{eq:magnet_analyt} in the compact form: 
\begin{equation}
    \begin{split}
& \mean{M(t+1)-M(t)}= 2\Bigg\{\sum_{n=3}^N\pi(n)(n-1)\times  \\ & \times \frac{1-M^2(t)}{4N}\left[F\big(\hat{k},\hat{n},\hat{p}_{-}\big)- F\big(\hat{k},\hat{n},\hat{p}_{+}\big)\right]\Bigg\}
    \end{split}
    \label{eq:magnet_analyt_simp}
\end{equation}
where $F(\hat{k},\hat{n},\hat{p})=Pr(X\leq \hat{k})=\sum_{i=0}^{\hat{k}}\binom{\hat{n}}{i}\hat{p}^i(1-\hat{p})^{\hat{n}-i}$ is the cumulative distribution function of the binomial distribution with $\hat{k},\hat{n}\in\mathbb{Z}$ and $\hat{p}\in[0,1]$.

\begin{figure}[htbp]
    \centering
\includegraphics[width=0.4\textwidth]{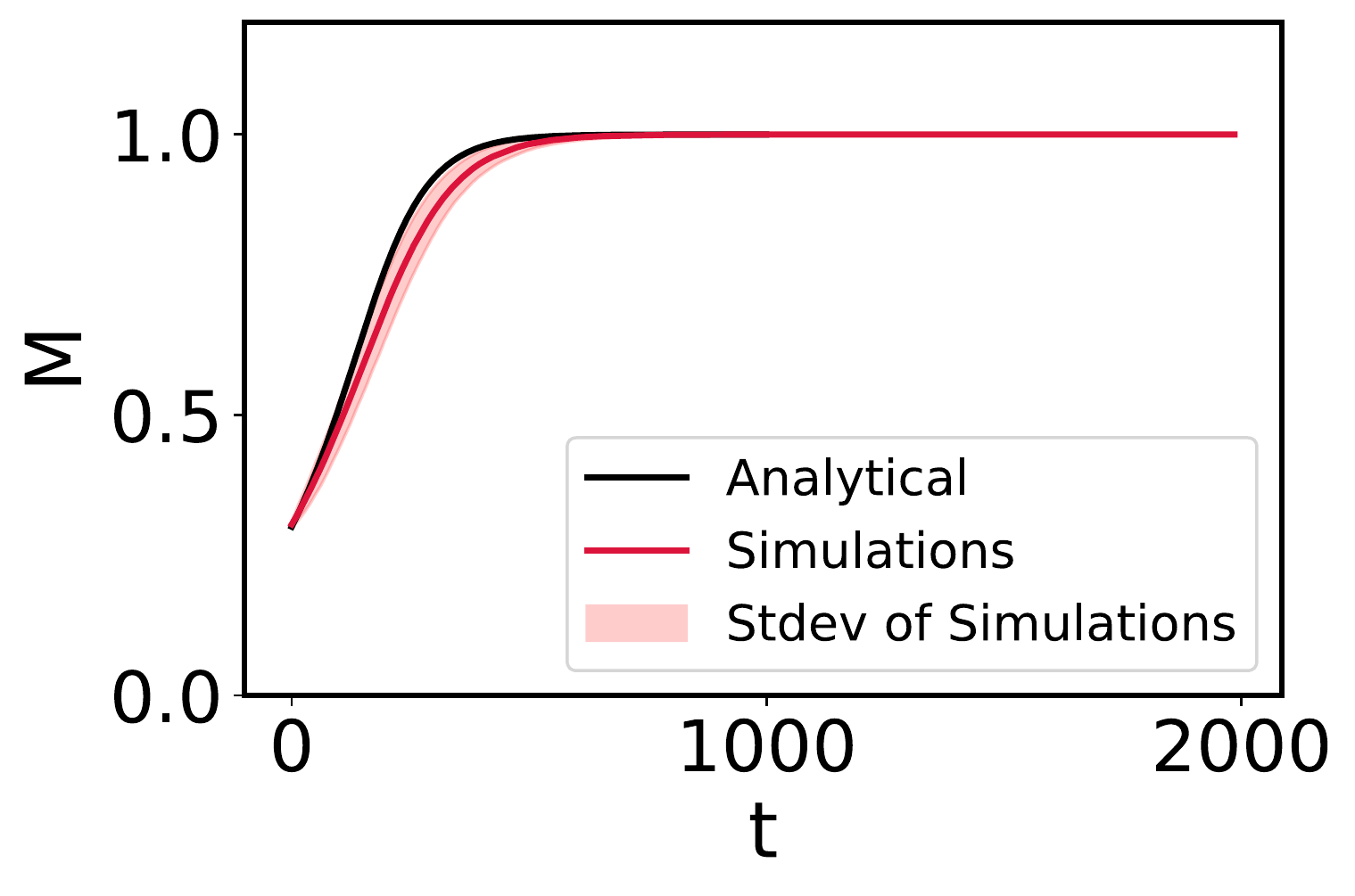}
    \caption{Dynamics of the average magnetization obtained from Eqn.~\eqref{eq:magnet_analyt_simp} (black line) and from simulations. Parameters: $N=500$, $\beta=3$, $\mean{M(0)}=0.3$, $r=0.5$, $\gamma=0.3$. The red points and the shaded area are the average and standard deviation of the average final magnetization for $20$ initial configurations respectively. For each configuration, we simulate $50$ trajectories.}
    \label{fig:comp_sequences}
\end{figure}

In Figure~\ref{fig:comp_sequences} we compare this analytical solution with computer simulations for a case where the average magnetization reaches $1$.
There is a good fit between the curves with an approximate  accuracy of $88\%$.

The sum in Eqn.~\eqref{eq:magnet_analyt_simp} converges for  $N\rightarrow \infty$.
This is because the series converges for $\hat{n} \rightarrow \infty$ due to the fact that the bound of $F(\hat{k},\hat{n},\hat{p})$ decays exponentially as $\hat{n} \rightarrow \infty$ based on Hoeffding's inequality. 
This convergence was also used to express the gradient of the potential $U(\mean{M})$, Eqn.~\eqref{dmdt_noise}, independently of finite size effects. 

It is interesting to note that the factors depending on the probability of rewiring $r$, Eqs.~\eqref{eq:w_0_to_1_final}, \eqref{eq:w_1_to_0_final}, cancel out in Eqn.~\eqref{eq:magnet_analyt_simp}. 
That means the dynamics of the average magnetization does not depend on the pairwise interactions expressed by $n=2$.

This reminds on the  adaptive voter model without group interactions
\citep{horstmeyer2020adaptive}, where the average magnetization of the system is conserved, i.e. it depends only on the initial magnetization, but not on the system size or on the probability of rewiring, $r$ \citep{horstmeyer2020adaptive}.

We can cover the behavior of the adaptive voter model without group interactions by setting $\beta=0$.
Notably, the group interactions in our model lead to a stochastic drift of the average magnetization.
Therefore the average magnetization is \emph{not} conserved.
The reason for the drift is that in our model the transition rates,  Eqs.~\eqref{eq:w_0_to_1_final}, \eqref{eq:w_1_to_0_final},  are nonlinear as opposed to the linear transition rates of the classical Voter Model.

Finally, for increasing values of $\gamma$, more terms in the inner sum of Eqs.~\eqref{eq:magnet_analyt}, \eqref{eq:magnet_analyt_simp} are added.
For increasing values of $\beta$ the values of $\pi(n)$ are increased for $n\geq 3$, which is at the expense of the $\pi(2)$ value. 
These effects accelerate in our model the dynamics of the average magnetization.

\end{appendix}
\end{document}